\documentclass[pra,preprint,showpacs,aps,amsmath,amssymb]{revtex4-1}
\usepackage{epsf}
\usepackage{color}
\usepackage{dcolumn}% Align table columns on decimal point
\usepackage{bm}% bold math
\everymath{\displaystyle}

\begin{document} %%%%%%%%%%%%%%%%%%%%%%%%%%%%%%%%%%%%%%%%%%%%%%%%%

\title{Generalization of exact operators of the Foldy-Wouthuysen transformation to arbitrary-spin particles in nonstationary fields} %Eriksen and exponential

\author{Alexander J. Silenko}
\email{alsilenko@mail.ru} \affiliation{Bogoliubov Laboratory of Theoretical Physics, Joint Institute for Nuclear Research, Dubna 141980, Russia}

%\date{file ``ostor12.tex", \today}

\begin {abstract}
Time-dependent terms in Hamiltonians and equations of motion are rather important for a quantum-mechanical description of particles with arbitrary spins in nonstationary fields. We use the Foldy-Wouthuysen representation which allows one to obtain the Schr\"{o}dinger picture of relativistic quantum mechanics. We determine exact nonexponential and exponential operators of the Foldy-Wouthuysen transformation for arbitrary-spins particles in the nonstationary case. Some fundamental properties of these operators and the Foldy-Wouthuysen Hamiltonian are also identified.
\end{abstract}
%\pacs {03.65.-w, 11.10.Ef} 
\maketitle

\section{Introduction}\label{Introduction}

The seminal Foldy-Wouthuysen (FW) representation discovered by Foldy and Wouthuysen 75 years ago \cite{FW} has unique and wonderful properties (see Ref. \cite{PRAFW} and references therein). The Hamiltonian and all operators are even, i.e., block-diagonal (diagonal in two spinors or spinor-like wave functions). The FW representation is equivalent to the Schr\"{o}dinger (Schr\"{o}dinger-Pauli) one in nonrelativistic quantum mechanics (QM). As a result, the form of quantum-mechanical operators for \emph{relativistic} particles
in external fields is usually the same as in the nonrelativistic QM. In particular, the position (Newton-Wigner)
operator \cite{NW} and the momentum one are equal to $\bm r$ and $\bm p=-i\hbar\nabla$, respectively (see Refs. \cite{Reply2019,PRAFW} for more details). The passage to the classical limit usually reduces to a replacement of the operators in quantum-mechanical Hamiltonians and equations of motion
with the corresponding classical quantities \cite{PRAFW,JINRLett12}. The probabilistic interpretation of wave functions lost in the Dirac representation is restored in the FW one \cite{PRAFW,Reply2019}. Thanks to these unique properties, the FW representation is now widely used not only in physics but also in quantum chemistry. In the latter case, it is reached with computer calculations.

Many transformation methods allowing one to derive a block-diagonal
Hamiltonian do not lead to the FW representation
(see Refs. \cite{erik,dVFor,PhysRevD,JMPcond,PRA2016}). The transformation operator can be presented in the exponential form. It has been proven in Refs. \cite{E,erik} that
%the block-diagonalization of an initial Hamiltonian \emph{always} transforms
%it to the FW representation if
the \emph{resulting} exponential operator of the FW transformation
should be odd and Hermitian. Paradoxically, the original method by Foldy and Wouthuysen \cite{FW}
does not satisfy this
requirement and does not lead to the FW representation
\cite{erik,dVFor,PRA2016}. In Ref. \cite{PRA2016}, main distinctive features of the FW
transformation have been considered and a possibility to correct the original method by Foldy and Wouthuysen has been shown. Fundamental properties of the FW transformation operators are actively used in relativistic quantum chemistry \cite{Dyall,ReiherWolfBook}.

The exact operator of the FW transformation has been obtained by Eriksen \cite{E}. The exact exponential FW transformation operator has been derived in Ref. \cite{PRAExpO}. The both operators are applicable only in the stationary case. The derivation of the corresponding operators for particles with an arbitrary spin in nonstationary fields is the goal of the present study.

We use the system of units $\hbar=1,~c=1$ but include $\hbar$ and $c$
explicitly when this inclusion
clarifies the problem.

%The paper is organized as follows. Previously obtained results related to the topic of the present study are considered in Sec. \ref{Previously}. The appropriate operator extraction of a square root in the operator of the exact FW transformation is fulfilled in Sec. \ref{Extraction}. In Sec. \ref{FWH}, we derive the relativistic FW Hamiltonian with the leading correction in the weak-field approximation. The corresponding relativistic wave equation of the second order is found in the general form in Sec. \ref{SecondOrderE}. A Dirac particle in an electrostatic field is considered as an example in Sec. \ref{Example}. The results obtained are discussed and summarized in Sec. \ref{Discussion}.

\section{Known operators of the exact Foldy-Wouthuysen transformation and a possibility of their generalization} \label{Previously}

For Dirac particles, one usually presents the initial Hamiltonian in the form
\begin{equation} {\cal H}_D=\beta m+{\cal E}+{\cal
O},\qquad\beta{\cal E}={\cal E}\beta,
\qquad\beta{\cal O}=-{\cal O}\beta. \label{eq3Dirac} \end{equation} 
The Dirac matrix $\beta$ commutes and anticommutes with the even and odd operators ${\cal E}$ and ${\cal O}$, respectively.

For a particle with an arbitrary spin, the initial Hamiltonian can be written as follows: 
\begin{equation} {\cal H}=\beta {\cal M}+{\cal E}+{\cal
O},\qquad\beta{\cal M}={\cal M}\beta,\qquad\beta{\cal E}={\cal E}\beta,
\qquad\beta{\cal O}=-{\cal O}\beta. \label{eq3G} \end{equation} The first term in this Hamiltonian contains the even operator.

Dirac operators act on a bispinor wave function. It has the form $\Psi=\left(\begin{array}{c} \phi \\ \chi \end{array}\right)$, where $\phi$ and $\chi$ are upper and lower spinors. Similarly, Hamiltonians and wave functions of particles with any spin can be written down in the bispinor-like form and can be defined by the same formula. In this case, $\phi$ and $\chi$ have $2s+1$ components, where $s$ is the spin quantum number. The matrix $\beta$ is block-diagonal and is defined by $\beta=\left(\begin{array}{cc} {1}& 0 \\ 0& -{1} \end{array}\right)$, where $-1,~0,~1$ are the corresponding $(2s+1)\times(2s+1)$ matrices.

In the general case, a transformation to a new representation
described by the wave function $\Psi'$ is performed with the
unitary operator $U$:
\begin{equation} \Psi'=U\Psi=\exp{(iS)}\Psi, \label{eqUiS} \end{equation} where $S$ is an exponential transformation operator.

Main properties of the \emph{exact} operators of the FW transformation, $S_{FW}$ and $U_{FW}=\exp{(iS_{FW})}$, have been determined by Eriksen \cite{E} and have been substantiated by Eriksen and Kolsrud
\cite{erik}. The FW transformation is \emph{unique} if the
operator $S_{FW}$ is \emph{odd},
\begin{equation} \beta S_{FW}=-S_{FW}\beta,
\label{VveEfrt} \end{equation} and Hermitian
($\beta$-pseudo-Hermitian for bosons \cite{PRA2016}). Some additional explanation of the
Eriksen method is given in Ref. \cite{PRA2016}.

The condition (\ref{VveEfrt}) is equivalent to \cite{E,erik}
\begin{equation} \beta U_{FW}=U^\dag_{FW}\beta.\label{Erikcon} \end{equation}

Thus, the FW transformation operator should satisfy Eq.
(\ref{Erikcon}) and should perform the transformation in one step.
Eriksen \cite{E} has found an operator possessing these
properties. To determine its explicit form \emph{in the stationary case}, one can introduce the sign
operator $\lambda={\cal H}/({\cal H}^2)^{1/2}$. The operator $({\cal H}^2)^{1/2}$ should be even if ${\cal E}=0$. To unambiguously define the square root, this statement should be complemented by the condition that the square root of the unit matrix ${\cal I}$ is equal to the unit matrix \cite{JMP}. The operator $1+\beta\lambda$ cancels either lower or upper
spinor for positive and negative energy states, respectively. The numerator and denominator of the operator $\lambda$ commute and \cite{E}
\begin{equation}\lambda^2=1, \quad [\beta\lambda,\lambda\beta]=0, \quad [\beta,(\beta\lambda+\lambda\beta)]=0.\label{eq3X3}
\end{equation}
Therefore, the operator of the exact FW transformation has the form \cite{E}
\begin{equation}
U_{FW}=\frac{1+\beta\lambda}{\sqrt{2+\beta\lambda+\lambda\beta}},
\qquad \lambda=\frac{{\cal H}}{({\cal H}^2)^{1/2}}. \label{eqXXI}
\end{equation} This equation and the definition of the square root remain unchanged for a particle with an arbitrary spin. 
The operator $({\cal H}^2)^{1/2}$ should be even if ${\cal M}=const$ and ${\cal E}=0$. The initial Hamiltonian operator, ${\cal H}$, is arbitrary. The numerator and denominator of the operator $U_{FW}$ commute. The even operator $\beta\lambda+\lambda\beta$ acting on the wave function with a single nonzero
spinor does not make another spinor be nonzero.

The equivalent form of the operator $U_{FW}$ \cite{JMPcond} shows that it is properly unitary ($\beta$-pseudounitary for bosons):
\begin{equation}
U_{FW}=\frac{1+\beta\lambda}{\sqrt{(1+\beta\lambda)^\dag(1+\beta\lambda)}}.
\label{JMP2009}
\end{equation}

This equation can also be used for a particle with
an arbitrary spin. In this case, the initial Hamiltonian is given by Eq.
(\ref{eq3G}). An exact exponential FW transformation operator \emph{in the stationary case} has been found in Ref. \cite{PRAExpO}. In the present study, we generalize the Eriksen and exponential FW transformation operators to the nonstationary case.

In this case, any unitary transformation involves not only the Hamiltonian operator but also the $-i\frac{\partial}{\partial
t}$ one. As a result, the Hamiltonian operator in the new representation takes the form
\begin{equation} {\cal H}'=U\left({\cal H}-i\hbar\frac{\partial}{\partial
t}\right)U^{-1}+ i\hbar\frac{\partial}{\partial t}  \label{taeq2} \end{equation} or
\begin{equation} {\cal H}'=U{\cal H}U^{-1}-i\hbar U\frac{
\partial U^{-1}}{\partial t}. \label{eq2}
\end{equation}

Equation (\ref{taeq2}) can be written as follows:
\begin{equation} {\cal H}'-i\hbar\frac{\partial}{\partial
t}=U\left({\cal H}-i\hbar\frac{\partial}{\partial
t}\right)U^{-1}.
%=U\left(\beta{\cal M}+{\cal E}+{\cal O}-i\hbar\frac{\partial}{\partial t}\right)U^{-1}.
\label{taeq3}
\end{equation} An inclusion of the time derivative in the process of transformation differs the Hamiltonian from other operators (see Ref. \cite{PRAnonstat}).

Equation (\ref{taeq2}) can be presented in the form
\begin{equation} {\cal H}'-i\hbar\frac{\partial}{\partial
t}=U\left({\cal H}-i\hbar\frac{\partial}{\partial
t}\right)U^{-1}=U\left(\beta{\cal M}+{\cal E}+{\cal
O}-i\hbar\frac{\partial}{\partial
t}\right)U^{-1}.
\label{taeq7}
\end{equation}
Equation (\ref{taeq7}) expresses a rather important property of the FW transformation for a particle in nonstationary (time-dependent) fields \cite{PRA2016,preprin}. %Evidently, one transforms the operator ${\cal H}-i\frac{\partial}{\partial
% t}$ but not the Hamiltonian operator alone.
Transformations of two even operators, ${\cal E}$ and
$-i\hbar\frac{\partial}{\partial t}$, are very similar. As a
result, the FW Hamiltonian (except for terms without commutators)
contains these operators only in the combination \begin{equation}{\cal F}={\cal
E}-i\hbar\frac{\partial}{\partial t}.\label{eqf}
\end{equation} Therefore, a transition
from a stationary to a nonstationary case can be performed with a
replacement of ${\cal E}$ with ${\cal F}$ in all terms containing
commutators \cite{PRA2016,preprin,TMPFW}. This important property allows us to solve the stated problem.

\section{General form of exact Foldy-Wouthuysen transformation operators} \label{Three}

It is easy to check that the Eriksen operator (\ref{eqXXI}) is not applicable in the nonstationary case. 
%The results obtained in Refs. \cite{E,erik,PRAExpO} are applicable only in the stationary case. 
In the stationary case, the operators ${\cal H}$ and $\frac{\partial}{\partial t}$ commute $\left(\frac{\partial{\cal H}}{\partial t}=0\right)$. As a result,
\begin{equation} 
U_{FW}\left({\cal H}-i\hbar\frac{\partial}{\partial
t}\right)U^{-1}_{FW}=U_{FW}{\cal H}U^{-1}_{FW}-i\hbar\frac{\partial}{\partial
t}={\cal H}_{FW}-i\hbar\frac{\partial}{\partial
t}.
\label{eqk}
\end{equation}
Since $[{\cal H},({\cal H}^2)^{1/2}]=0$ and ${\cal H}+\beta{\cal H}\beta=2(\beta {\cal M}+{\cal E})$, we obtain the following formula:
\begin{equation}\begin{array}{c} 
U_{FW}{\cal H}U^{-1}_{FW}=\frac{1}{\sqrt{2+\beta\lambda+\lambda\beta}}\left(1+\beta\lambda\right){\cal H}\left(1+\lambda\beta\right)\frac{1}{\sqrt{2+\beta\lambda+\lambda\beta}}\\=\frac{1}{\sqrt{2+\beta\lambda+\lambda\beta}}\left[2(\beta {\cal M}+{\cal E})+\beta\sqrt{{\cal H}^2}+\sqrt{{\cal H}^2}\beta\right]\frac{1}{\sqrt{2+\beta\lambda+\lambda\beta}}.
\end{array} \label{eqken}
\end{equation}
Because the anticommutator $\{\beta,A\}\equiv\beta A+A\beta$ is even for any matrix operator $A$, the operator (\ref{eqken}) is even. In the nonstationary case, the transformation of the operator $-i\hbar\frac{\partial}{\partial t}$ is nontrivial. It is easy to obtain that
\begin{equation}\begin{array}{c} 
U_{FW}\frac{\partial}{\partial t}U^{-1}_{FW}=\frac{1}{\sqrt{2+\beta\lambda+\lambda\beta}}\left(1+\beta\lambda\right)\frac{\partial}{\partial t}\left(1+\lambda\beta\right)\frac{1}{\sqrt{2+\beta\lambda+\lambda\beta}},\\ \left(1+\beta\lambda\right)\frac{\partial}{\partial t}\left(1+\lambda\beta\right)=3\frac{\partial}{\partial t}+
\beta\lambda\frac{\partial}{\partial t}+\lambda\frac{\partial}{\partial t}\beta+\dot{\lambda}\beta+\frac12\beta[\dot{\lambda},\lambda]\beta.
\end{array} \label{eqnst}
\end{equation}
The operators $\dot{\lambda}\beta$ and $\frac12\beta[\dot{\lambda},\lambda]\beta$ are not even and the sum of the three precedent operators is even. Therefore, the operator $U_{FW}\left({\cal H}-i\hbar\frac{\partial}{\partial
t}\right)U^{-1}_{FW}$ is not even and the FW transformation operator (\ref{eqXXI}) obtained by Eriksen for the stationary case become inapplicable in the nonstationary case.

We can identify the general form of the FW transformation operator applicable in the both cases. Section \ref{Previously} shows that we need to replace the operator $\lambda$ in Eqs. (\ref{eqXXI}) and (\ref{JMP2009}) with the new operator
\begin{equation}
\Lambda=\frac{{\cal H}-i\hbar\frac{\partial}{\partial
t}}{\left[\left({\cal H}-i\hbar\frac{\partial}{\partial
t}\right)^2\right]^{1/2}}=\frac{\beta {\cal M}+{\cal F}+{\cal
O}}{\left[\left(\beta {\cal M}+{\cal F}+{\cal
O}\right)^2\right]^{1/2}}.
\label{eqLam}
\end{equation}
As a result,
\begin{equation}
U_{FW}=\frac{1+\beta\Lambda}{\sqrt{2+\beta\Lambda+\Lambda\beta}}=\frac{1+\beta\Lambda}{\sqrt{(1+\beta\Lambda)^\dag(1+\beta\Lambda)}}. \label{enXXI}
\end{equation}  Since ${\cal H}\Psi=i\hbar\frac{\partial\Psi}{\partial t}=E(t)\Psi$, where $E(t)$ is the time-dependent energy, the operator $1+\beta\Lambda$ cancels either lower or upper spinor for positive- and negative-energy states, respectively.
The FW Hamiltonian is given by [cf. Eq. (\ref{eqken})]
\begin{equation}\begin{array}{c} 
{\cal H}_{FW}=U_{FW}\left({\cal H}-i\hbar\frac{\partial}{\partial t}\right)U^{-1}_{FW}+i\hbar\frac{\partial}{\partial t}\\=\frac{1}{\sqrt{2+\beta\Lambda+\Lambda\beta}}\left[2(\beta {\cal M}+{\cal F})+\beta\sqrt{{\cal H}^2}+\sqrt{{\cal H}^2}\beta\right]\frac{1}{\sqrt{2+\beta\Lambda+\Lambda\beta}}+i\hbar\frac{\partial}{\partial t}.
\end{array} \label{eqkenFW}
\end{equation} Evidently, this Hamiltonian is even. 

The simplest method to determine the exact exponential operator of the FW transformation in the nonstationary case is based on the results obtained in Ref. \cite{E}.  As follows from Eqs. (\ref{eqUiS}), (\ref{eqXXI}), and (\ref{eqf}),
\begin{equation}\begin{array}{c} 
\sin{S_{FW}}=-\frac{i}{2}\left(U_{FW}-U_{FW}^{-1}\right)=-i\frac{\beta\Lambda-\Lambda\beta}{2\sqrt{2+\beta\Lambda+\Lambda\beta}},\\ 
\cos{S_{FW}}=\frac12\left(U_{FW}+U_{FW}^{-1}\right)=\frac{\sqrt{2+\beta\Lambda+\Lambda\beta}}{2}.
\end{array} \label{eqUexpn} \end{equation}
Therefore (cf. Ref. \cite{E}), 
\begin{equation}\begin{array}{c} 
\sin{2S_{FW}}=2\sin{S_{FW}}\cos{S_{FW}}=-\frac{i}{2}\left(\beta\Lambda-\Lambda\beta\right),\\ \cos{2S_{FW}}=2\cos^2{S_{FW}}-1=\frac{1}{2}\left(\beta\Lambda+\Lambda\beta\right).
\end{array} \label{eqtUexp} \end{equation}
As follows from Eq. (\ref{eqtUexp}), in the nonstationary case the exponential operator is defined by the two equivalent relations:
\begin{equation}\begin{array}{c} 
S_{FW}=-\arcsin{\frac{i\left(\beta\Lambda-\Lambda\beta\right)}{2\sqrt{2+\beta\Lambda+\Lambda\beta}}},\\ S_{FW}=-\frac12\arcsin{\frac{i\left(\beta\Lambda-\Lambda\beta\right)}{2}}.
\end{array} \label{eqttexp} \end{equation}
The well-known expansion of arcsine into a series shows that the latter relation can be written in the form (cf. Ref. \cite{PRAExpO})
\begin{equation}
S_{FW}=-\frac{\beta}{2}\arcsin{\frac{i\left(\Lambda-\beta\Lambda\beta\right)}{2}}.
\label{eqtlexp} \end{equation}
Equations (\ref{eqttexp}) and (\ref{eqtlexp}) solve the problem of the general form of the exact exponential FW transformation operator.

\section{Fundamental properties of the general exact Foldy-Wouthuysen transformation operators and the Foldy-Wouthuysen Hamiltonian} \label{FundamentalP}

We can determine some fundamental properties of the general exact FW transformation operators and the FW Hamiltonian. The FW Hamiltonian should be block-diagonal but this property is necessary but not sufficient \cite{JMPcond,dVFor,PhysRevD}. Certainly, the obtained general exact FW transformation operators satisfies the Eriksen conditions. The exponential
operator $S_{FW}$ defined by Eqs. (\ref{eqttexp}), (\ref{eqtlexp}) is odd and Hermitian ($\beta$-pseudo-Hermitian for bosons). The nonexponential operator $U_{FW}$ satisfies the property (\ref{Erikcon}). 

We can additionally indicate other fundamental properties of the operators $S_{FW}$ and $U_{FW}$. It has been proven in Ref. \cite{Validation} that the Eriksen operator $\lambda$ does not depend on ${\cal E}$ when $[{\cal O},{\cal E}]=0$. We can generalize this property and apply it in the nonstationary case. First of all, we should present some important relations: 
\begin{equation}
\frac{\partial \lambda}{\partial{\cal E}}=0,
\label{eqfirst} \end{equation}
\begin{equation}\frac{\partial}{\partial{\cal E}}[{\cal O},{\cal E}]=0,\qquad \frac{\partial}{\partial{\cal E}}\{{\cal O},{\cal E}\}=2{\cal O},\qquad \frac{\partial}{\partial{\cal E}}[{\cal O},{\cal E}^2]=2[{\cal O},{\cal E}].\label{eqecond} \end{equation}
Evidently, Eq. (\ref{eqfirst}) should be valid when $[{\cal O},{\cal E}]\neq0$. Thus, the relations (\ref{eqfirst}) and  (\ref{eqecond}) explicitly show that any expansion of $\lambda$ in a series (see Refs. \cite{dVFor,TMPFW,VJ,preprin}) can contain terms proportional to $[{\cal O},{\cal E}]$ but not terms proportional to $[{\cal O},{\cal E}^2]=\{[{\cal O},{\cal E}],{\cal E}\}$ or $\{{\cal O},{\cal E}\}[{\cal O},{\cal E}]$. The validity of any other terms can be properly checked.

We can now pass to the general case of a particle with any spin in a nonstationary field. As follows from Eq. (\ref{eqLam}),
\begin{equation}
\frac{\partial \Lambda}{\partial{\cal F}}=0.
\label{eqfiLat} \end{equation}
Therefore, any expansion of $\Lambda$ in a series can contain terms proportional to $[(\beta{\cal M}+{\cal O}),{\cal F}]$ but not terms proportional to $[(\beta{\cal M}+{\cal O}),{\cal F}^2]=\{[(\beta{\cal M}+{\cal O}),{\cal F}],{\cal F}\}$ or $\{(\beta{\cal M}+{\cal O}),{\cal F}\}[(\beta{\cal M}+{\cal O}),{\cal F}]$. As follows from Eqs. (\ref{eq3G}), (\ref{enXXI}), (\ref{eqttexp}), (\ref{eqtlexp}), and (\ref{eqfiLat}), 
\begin{equation} \begin{array}{c}
\frac{\partial U_{FW}}{\partial{\cal F}}=\frac{\partial U^{-1}_{FW}}{\partial{\cal F}}=0,\qquad \frac{\partial S_{FW}}{\partial{\cal F}}=\frac{\partial S^{-1}_{FW}}{\partial{\cal F}}=0,\\ \frac{\partial}{\partial{\cal F}}\left[U_{FW}\left({\cal H}-i\hbar\frac{\partial}{\partial t}\right)U^{-1}_{FW}\right]=\frac{\partial}{\partial{\cal F}}\left({\cal H}_{FW}-i\hbar\frac{\partial}{\partial t}\right)=1.
\end{array} \label{eqfiLit} \end{equation}
Evidently,
\begin{equation}
\frac{\partial}{\partial{\cal F}}\left({\cal H}_{FW}-{\cal E}\right)=\frac{\partial}{\partial{\cal F}}\left({\cal H}_{FW}-i\hbar\frac{\partial}{\partial t}-{\cal F}\right)=0.
\label{eqfiLfn} \end{equation}

The derived equations define some fundamental properties of the general exact FW transformation operators and the FW Hamiltonian. All correct approximate FW transformation operators and Hamiltonians should satisfy the properties (\ref{eqfiLat}) -- (\ref{eqfiLfn}). In particular, these properties are met for the FW Hamiltonian obtained in Refs. \cite{VJ,PRA2016,preprin,TMPFW} as a power series in ${\cal E}/m$ and ${\cal O}/m$.

\section{Discussion and summary} \label{Discussion}

The exact FW transformation operator (\ref{eqXXI}) has been found by Eriksen \cite{E} in the stationary case for a spin-1/2 particle. Wonderfully, this operator can be used without any changes for a particle with an arbitrary spin. However, it does not cover the case of nonstationary external fields. This problem is solved in the present paper. Unfortunately, neither the derived operators (\ref{enXXI}), (\ref{eqttexp}), and (\ref{eqtlexp}) nor other methods (see Refs. \cite{PRA2016,Dyall,ReiherWolfBook,preprin} and references therein) allow one to determine an explicit form of a FW Hamiltonian. Relativistic quantum chemistry (as well as relativistic quantum mechanics) uses numerous methods satisfying the Eriksen conditions and basing on an expansion of a final FW Hamiltonian in a series in powers of the potential divided by the total kinetic energy $\biggl(\frac{V}{\sqrt{m^2c^4+c^2p^2}}$ or $\biggl\{{\cal E},\frac{1}{\sqrt{{\cal M}^2+{\cal O}^2}}\biggr\}\biggr)$. In Ref. \cite{preprin}, a similar approach has been based on the exact Eriksen operator (\ref{eqXXI}). In this paper, the weak-field approximation resulting in $|{\cal E}|\ll\sqrt{m^2+{\cal O}^2}$ has been used for a spin-1/2 particle.

Our analysis shows that the connection between the exact FW transformation operators in the stationary and nonstationary cases results in the replacement of ${\cal E}$ with ${\cal F}$. The present paper gives an important and completely rigorous proof of this replacement (see Sec. \ref{Three}). The general exact FW transformation operator cancels either lower or upper spinor for positive- and negative-energy states, respectively. The obtained FW Hamiltonian (Hamiltonian in the FW representation) is even. These properties are very important and conclusively prove the validity of the substitution ${\cal E}\rightarrow{\cal F}$ not only into the final FW Hamiltonian but also into the Eriksen operator (\ref{eqXXI}). 

The relativistic FW Hamiltonian which has been derived in Ref. \cite{preprin} in the weak-field approximation for a spin-1/2 particle (${\cal H}={\cal H}_{D}$) is given by
\begin{equation}
\begin{array}{c}
{\cal H}_{FW}=\beta\epsilon+ {\cal E}-\frac 18\left\{\frac{1}
{\epsilon(\epsilon+m)},[{\cal O},[{\cal O},{\cal
F}]]\right\}+\frac{1}{64}\left\{\frac{2\epsilon^2-m^2}{\epsilon^4(\epsilon+m)^2},[{\cal O}^2,[{\cal O}^2,{\cal
F}]]\right\},\\ \epsilon=\sqrt{m^2+{\cal O}^2}.
\end{array}
\label{Hamwfin}
\end{equation}
In Ref. \cite{preprin}, the substitution ${\cal E}\rightarrow{\cal F}$ has been fultilled into the final FW Hamiltonian and has been substantiated with Eq. (\ref{taeq7}). The used approach is applicable for a derivation of corrections quadratic (bilinear) in ${\cal E}$ for a spin-1/2 particle in nonstationary external fields \cite{preprin}. This possibility is rather promising because the FW representation gives one a clear physical interpretation of relativistic Hamiltonians. On the other side, taking into account corrections quadratic in ${\cal E}$ usually covers a precision needed in nonrelativistic quantum electrodynamics (see Ref. \cite{preprin} and references therein) and for the solution of the strong-field ionization problem (see Ref. \cite{comnonstat}). We underline that the description of particles in nonstationary fields in the FW representation plays an important role in contemporary physics.

In summary, we have generalized the previously obtained results \cite{E,PRAExpO} and have derived the exact nonexponential and exponential operators of the FW transformation for arbitrary-spin particles in nonstationary fields. The difference between new FW transformation operators and previously obtained ones reduces to the replacement of ${\cal E}$ with ${\cal F}={\cal
E}-i\hbar\frac{\partial}{\partial t}$ or, equivalently, ${\cal H}$ with ${\cal H}-i\hbar\frac{\partial}{\partial t}$ in all commutators. The present study justifies specific results obtained in Refs. \cite{PRA2016,preprin,TMPFW}. 
While an explicit expression for the FW Hamiltonian cannot be obtained, the derived operators can significantly simplify a calculation of relativistic FW Hamiltonians with a high accuracy. We have also determined some fundamental properties of the general exact Foldy-Wouthuysen transformation operators and the Foldy-Wouthuysen Hamiltonian.

\section*{Data availability statement}

All data that support the findings of this study are included within the article.

%\begin{acknowledgments}

%This work was supported in part by the Belarusian Republican Foundation for
%Fundamental Research (Grant No. $\Phi$14D-007) and by the Heisenberg-Landau Program of the German Ministry for Science and Technology (Bundesministerium f\"{u}r Bildung und Forschung).
%\end{acknowledgments}

\end{document}